# Heat flux deposition on plasma facing components using a convective model with ripple and Shafranov shift

**R. Mitteau** [a], A. Moal [b], J. Schlosser [a], D. Guilhem [a]

[a] *Association Euratom - CEA sur la Fusion Contrôlée. CEA Cadarache F-13108 Saint Paul Lez Durance*
[b] *CISI, Centre d'Etude de Cadarache F-13108 Saint Paul Lez Durance*

**Abstract**

A heat flux deposition code is a very useful tool for the design of plasma facing components. The classical cosine model representative of the convective parallel heat flux was coded within the standard finite element code of the CEA Castem 2000. Two perturbations to the idealised magnetic surfaces were taken into account: the Shafranov shift and the ripple of the toroidal magnetic field. The calculation scheme - named Tokaflu - was confronted to previous computations on bottom modular limiters and to experiments on the inner first wall. The code was used to optimise the shape of future plasma facing components in Tore Supra designed for the CIEL project, namely the toroidal pumped limiter and its neutralisers. Developments are under way to include the shadowing of the components, other power deposition schemes and possibly later plasma sections other than circular.

# 1. Introduction

Tokamak plasmas loose their energy by convection and radiation on first wall components. Convective losses are caused by the plasma particles which impinge on the surface of the components. The convective heat is spread in two components, one parallel to the magnetic field lines and the other perpendicular to the magnetic surfaces [1,2]. The parallel component raises the highest concern, since the heat flux is very high (>100 MW/m²). Such a value is one order of magnitude higher than allowed by today's technology (5-20 MW/m²). In order to sustain the heat flux, the surfaces of plasma facing components are inclined with respect to the magnetic lines to have small incident angles (<5°). Slightly higher incident angles increase the heat flux dramatically, which could lead to the deterioration of the components. Thus, the design of the first wall components requires modeling that are accurate enough to evaluate the incident angles and then the heat fluxes with sufficient precision. At Tore Supra, such modelling was already existing with partial codes devoted to the vertical modular limiters and the outboard limiter [3-5]. However the extensive design work for the "Tore Supra Continu" and then "CIEL" projects showed the necessity for an efficient and integrated software. A parallel convective heat deposition model - named Tokaflu - was developed and then coded within Castem 2000 which is the standard finite element code used by CEA for thermomechanical analysis [6]. The advantage of using a finite element code is that it is quite naturally based on geometrical objects. The geometrical description of the components can be as precise as desired. Plasma facing components have increasingly complex shapes to accommodate the high heat flux and the description of the component has to be realistic enough to avoid over simplification. Moreover, the heat flux pattern computed by the code can easily be used as input for thermal, mechanical or thermohydraulical computations. A similar procedure was followed for TEXTOR's limiter ALT-II [7].

# 2. Physical model for the convective parallel heat flux

The parallel convective heat follows the magnetic field lines. The first object met by the plasma defines a frontier: the last closed flux surface (LCFS). The heat flux along the field lines in the scrape of layer (SOL) decreases with the distance to the LCFS according to an exponential law with a characteristic e-folding length $\lambda_q$ (Fig. 1) [8]. On the surface of the plasma facing component, the incoming heat flux $\phi$ is a combination of the exponential decrease and of the angle between the magnetic field lines and the surface. It can be described by:

$$\phi = \phi_0 \cdot \exp\left(-\frac{\delta}{\lambda_q}\right) \cdot |\vec{b} \cdot \vec{n}| \quad (1)$$

where $\phi_0$ is the heat flux on the LCFS, $\delta$ the radial distance into the SOL to the LCFS, $\vec{b}$ the unit vector along the magnetic field and $\vec{n}$ the vector normal to the surface.

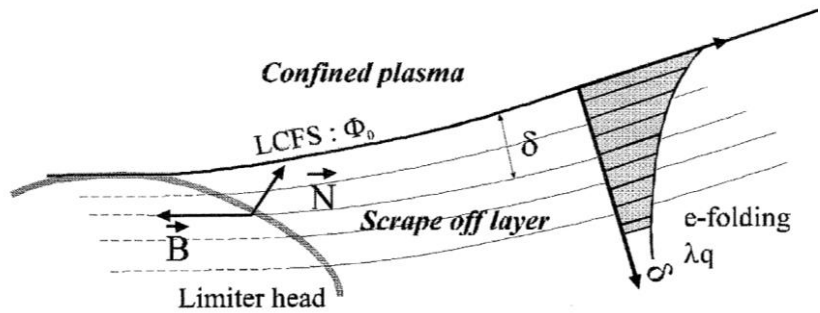

Fig. 1. Heat flux deposition principle.

The limitation of this model (so-called "cosine model") are known: on large area components, the perpendicular convective heat flux may not be neglected [1,2]. Moreover, some anomalous high heat flux is sometimes observed at the contact point to the plasma [9]. Therefore, in some cases, the parallel convective heat flux can't account for the majority of the heat deposition and the heat pattern can't be only approximated by the parallel convective heat flux (see also Section 4). The cosine model is however needed as the basic law governing heat deposition and is in some cases sufficient to describe the heat pattern.

In Tore Supra, two perturbations to the idealised geometry of perfectly toroidal nested surfaces with the same centres have been taken into account: the Shafranov shift and the ripple (which is the modulation of the toroidal magnetic field caused by the discrete coils).

The Shafranov shift is a change of the centres of the magnetic surfaces when their small radii changes. The centres of the magnetic surfaces shift towards the magnetic axis of the tokamak when their small radii increase. One of Tore Supra aims is to study advanced scenarii which have a high $\Lambda$ (equal to $\beta_p + \frac{l_i}{2} - 1$, $\beta_p$ being the poloidal beta and $l_i$ the internal plasma inductance) which causes high shift, so that it cannot be neglected. The formulas used in Tokaflu for the magnetic field come from Shafranov [10]:

$$B_{Pol}^{\rho}(M) = B_0 \cdot \frac{\rho_0}{2R_P} \left[ \left(1 - \frac{\rho_0^2}{(\rho_M^P)^2}\right) \left(\Lambda + \frac{1}{2}\right) + \ln\left(\frac{\rho_M^P}{\rho_0}\right) \right] \cdot \sin\theta_M^P \quad (2a)$$

$$B_{Pol}^{\theta}(M) = B_0 \cdot \frac{\rho_0}{\rho_M^P} + B_0 \cdot \frac{\rho_0}{2R_P} \left[ \left(1 + \frac{\rho_0^2}{(\rho_M^P)^2}\right) \left(\Lambda + \frac{1}{2}\right) + \ln\left(\frac{\rho_M^P}{\rho_0}\right) - 1 \right] \cdot \cos\theta_M^P \quad (2b)$$

where $B_{pol}$ is the poloidal magnetic field, $\theta$ refer to the poloidal angle and $\rho$ to the small radius, $B_0 = \frac{-\mu_0 I_p}{2\pi\rho_0}$, $\rho_M^P$ is the radius of the magnetic surface of M and $\theta_M^P$ its poloidal position, $\rho_0$ and $R_p$ are the small and major radii of the LCFS and $I_p$ the plasma current. These equations refer in fact to pseudo-toroidal reference axes defined at $R = R_p$.

Another effect of the Shafranov shift is to compress the magnetic surfaces on the low field side, hence diminishing the local $\lambda q$. The effect is opposite on the high field side.

The other perturbation taken into account is the ripple of the toroidal field. In Tore Supra, the ripple is not negligible because of the spacing between the 18 superconducting coils and because the plasma is close to the coils. A real account of this perturbation would require a magnetostatic computation at each run, expensive in CPU time. This would not be compatible with the performance awaited for the code, and also not realistic compared to the achievable accuracy. Therefore an empirical model was used [4,5,11]. It is based on exponential laws and describes the magnetic surfaces:

$$\rho_M^R = \rho_M^{mer} + \Delta\rho_M^{mer} \cdot \left[\cos\left(18 \cdot \varphi_M^R\right) - 1\right] \quad (3a)$$

with:

$$\Delta\rho_M^{mer} = a \cdot \exp\left(b \cdot \rho_M^{mer}\right) \cdot \exp\left(-c \; \theta_M^{R\,2}\right) \quad (3b)$$

φ is the toroidal angle, $\rho_M^{mer}$ is the radial coordinate of the magnetic surface in the meridian plane between two successive coils (φ = 0). $\Delta\rho_M^{mer}$ is the radial half-displacement of the field line between φ = 0 and φ = 10°. The Tore Supra coefficients a, b, c are parameters of the model which are determined from field lines computation [11]:

a = 5,8.10$^{-5}$ m ; b = 5,5 m$^{-1}$ ; c = 4,5.10$^{-5}$ (degree)$^{-2}$

The resulting magnetic surfaces with $\rho_M^{mer}$ = 0.75, 0.85 and 0.95 m are represented Fig. 2. The ripple effect is amplified 2.5 times. Λ is 0, causing a Shafranov shift which is observable on the figure.

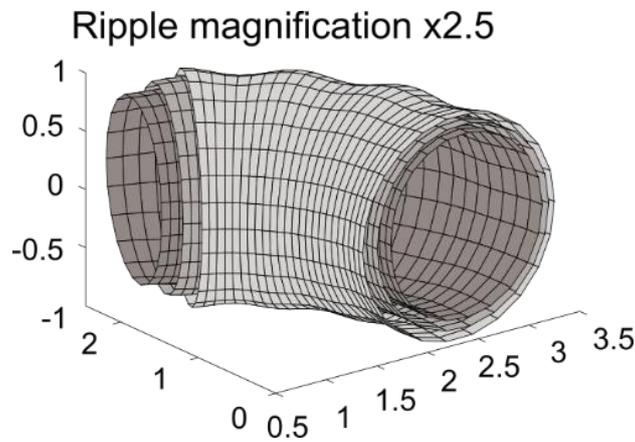

Fig. 2. Magnetic surfaces with ripple and Shafranov shift.

## 3. Numerical modelling

The model is developed in the finite element code Castem 2000 of CEA [6]. This code is object oriented, and possesses a macro language allowing to write subroutines. The mesh of the surface wetted by the plasma is created by using the standard tools of Castem 2000. The calculation scheme starts from that mesh. On the whole set of nodes, the subroutines compute:

a) the set of vector normal to the surface ;

b) the set of toroidal and poloidal fields according to section I ;

c) the set of distance to the LCFS, taking into account the real shape of the LCFS perturbed by the shift and the ripple ;

d) the set of the local $\lambda_q$, which varies according to the Shafranov shift. The changes of $\lambda_q$ (for parallel heat flux) caused by the ripple were neglected as a second order, which is shown in [5].

The relative value of the heat flux on the component is calculated according to Eq. (1). The actual heat flux needs the value of $\Phi_0$ on the LCFS. When the total parallel convective power is known, $\Phi_0$ is theoretically calculable by adjusting it so that integration over the whole surface of all wetted component sums up to the right power. Often, a main limiter removes the majority of the parallel convective power and a normalisation by integration of the heat flux profile on this limiter allows to compute the real heat flux value. On secondary limiters such as neutralisers or lateral protections for antennas, the $\Phi_0$ is obtained from previous values calculated on the main limiter.

The only input data needed by the code are the plasma current $I_p$, the plasma small and major radii a and $R_p$, the current in the toroidal coils $I_t$, $\Lambda$ (see above), the heat flux decay length for parallel heat flux $\lambda_q$ and the distance of the plasma center to the equatorial plane h. The coherence between all these parameters is however not considered, because it is typical of the tokamak and depends on the plasma scenario. The particles density isn't a direct parameter, its sole influence is through the heat flux decay length $\lambda_q$. In Tore Supra under certain conditions, these 7 parameters can be reduced to 6 by using the empirical law $\lambda_q = \frac{15.4}{\sqrt{I_p}}$, $I_p$ being expressed in mega amps and the result in millimetres [3].

The code runs in 3D, but computations can also be done in 2D section (poloidal section: toroidal angle $\varphi$ constant and $\rho,\theta$ variable. "Toroidal section": poloidal angle $\theta$ constant and $\varphi, \rho$ variable). This allows to perform easier calculations on smaller meshes.

A discussion on the precision of the code is difficult because of the numerous effects involved. Regarding the position of the field lines, the Shafranov shift can amount up to 18 millimetres for a typical advanced scenario with $\Lambda = 0.5$. The ripple is around 3 mm on the high field side but increases up to 30 mm on the low field side. The position of the components is known with a precision of 0.1 mm with the help of theodolites, but their displacement can reach 5 mm when the vessel is baked. The magnetic field lines are also known with a few millimetres errors due to small shifts in the coils positioning. The actual plasma ellipticity and triangularity can also introduce errors in the millimetre range. More important for the heat flux is the error made on the incidence angle, however even more difficult to evaluate.

## 4. Confrontation, validation and application of the code

Tokaflu was confronted to previous studies made on the bottom modular limiters [3]. Infrared imaging of these limiters was then correlated to 2D finite element computations. They were made with a customised subroutine, which included the ripple but not the Shafranov shift (negligible at this location). The correlation between experiments and computations was used to determine the heat flux decay length. Fig. 3 shows a good agreement between the previous computations (quoted as "reference") and the one made with the present code Tokaflu.

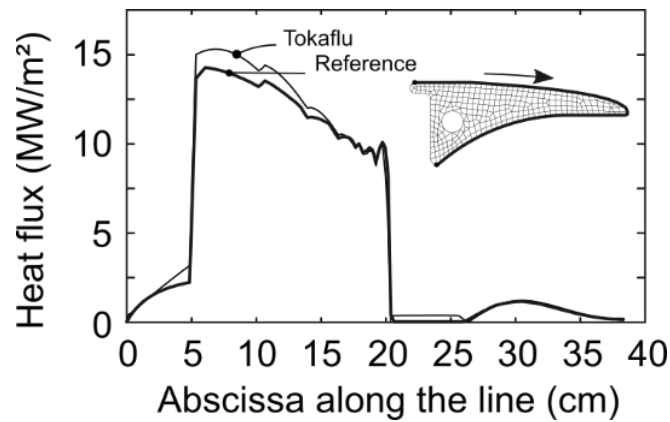

Fig. 3. Heat flux along the tiles of the modular bottom limiter.

An attempt was made to validate the code on the inner first wall, which is currently a fairly diagnosed plasma facing component in Tore Supra: infrared thermography, thermocouples, langmuir probes and calorimetry. The cosine model is however probably not best suited for the inner first wall, which may be subjected mainly to perpendicular heat flux [9,12]. Nevertheless, Fig. 4 show the temperature of thermocouples mounted on the compliance layer between the carbon fibre composite tiles and the stainless steel heat sink which are representative of the mean surface temperature over the tile, along with the profile of the heat flux calculated by Tokaflu. With actively cooled plasma facing components in steady-state, the temperature and the heat flux are proportional, and actually both profiles can be made overlapping. It shows that the cosine model when applied to the polygonal shape of the inner first wall gives a one-peak profile representative of the heat flux. This result had been already shown [13] but is another contribution to the validation of the code.

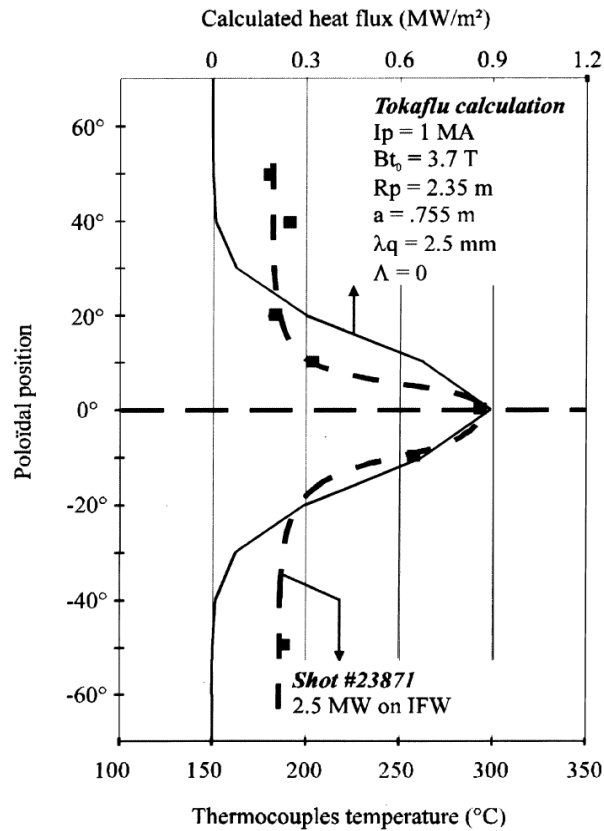

Fig. 4. Poloidal profiles of the measured temperature and of the calculated flux on the inner first wall of Tore Supra.

This code has been largely used in the design of the toroidal pumped limiter which is being fabricated in frame of the "CIEL" project [14]. This project aims to have long and stationary discharges in Tore Supra [15]. For this toroidal limiter, which has a relatively small poloidal extension and is located in the bottom of the machine where the code was confronted to the results on the modular limiters, the parallel convective model is suited. The parallel convective heat flux represents at least the highest constraint. The higher the fraction of perpendicular heat flux, the lower the fraction of parallel heat flux responsible for the peaking. Therefore the code is at least conservative. The map of the power deposition computed with Tokaflu is given Fig. 5. The leading edge receives 6 MW/m², with the profile being actually very peaked (the mean heat flux on the TPL is 2.2 MW/m²). On the flat part, large peaks up to 6 MW/m² are caused by the ripple at each side of the coil. These heat fluxes are compatible with the technology of active metal casting that was chosen for the toroidal pumped limiter. This calculation proved the viability of its concept.

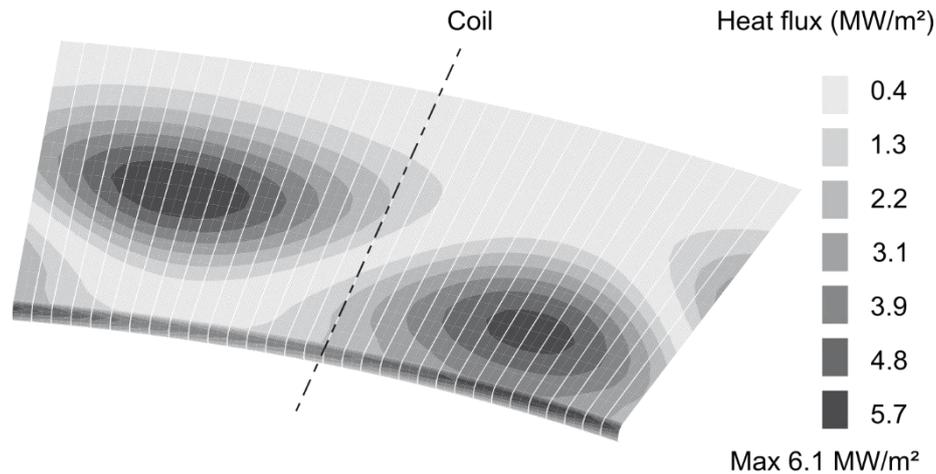

$Ip = 1.7$ MA / $\lambda q = 15$ mm / $Rp = 2.4$ m / $a = 0.72$ mm / $Rleading\ edge = 2.2$ m

Fig. 5. Heat flux on 20° of the toroidal pumped limiter.

Tokaflu was also used to optimise the shape of the neutraliser fingers located underneath the toroidal pumped limiter and also for the lateral protections for ICRH and lower hybrid antennas.

## 5. Future developments

Following developments are foreseen to fit more precisely the actual heat deposition pattern. Most important is the shadowing effect: studies on the inner first wall and on the toroidal pumped limiter showed that self shadowing may be an issue for these components. It affects only large area components which are overflied by the magnetic field lines over more than two periods of the ripple. The ripple causes field lines to enter and exit the component more than one time over a few coils (Fig. 6).

This effect tends to forbid the heat flux to reach certain areas of the limiter. Because of the normalisation of the $\Phi o$ to the total power removed by the component, this will increase the $\Phi o$ and thus the flux on the rest of the surface. It increases the peaking of the heat deposition. For that reason, the coding of this potentially negative effect is under way in Tokaflu. The possibility to take shadow effects into account will also be useful to work on the interaction between the components (main limiter/antennas protection or main limiter/neutralisers).

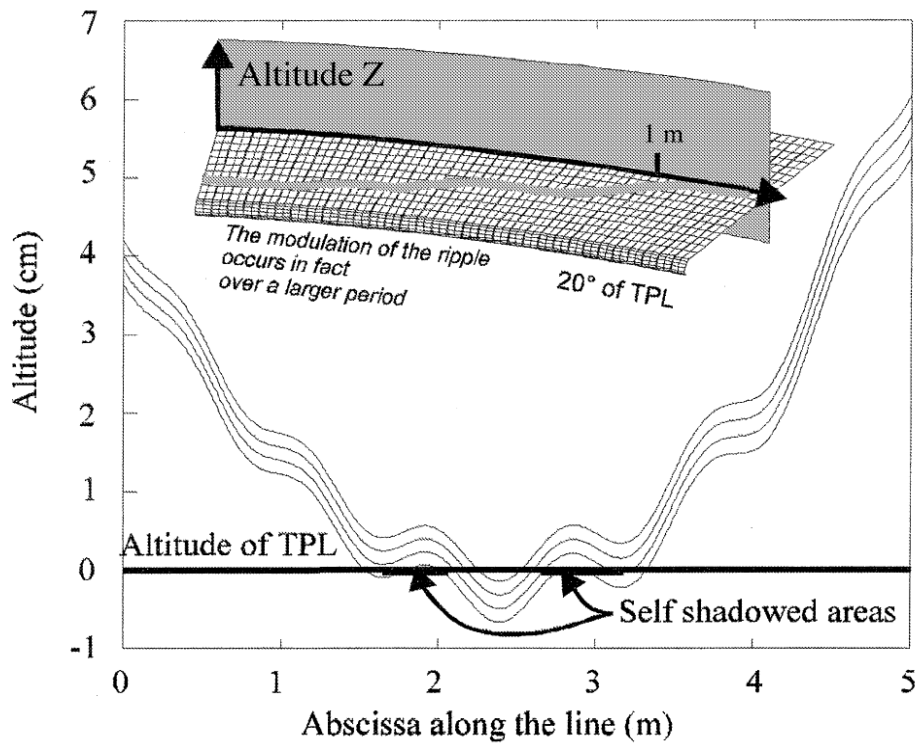

Fig. 6. Self shadowing due to ripple.

The model could also be improved by adding the contribution of the perpendicular and radiated heat fluxes to benefit from a more general model. The possibility to take into account the dissymmetry between ion and electron side is also envisioned. A further step would be to take into account the ellipticity and the triangularity of the plasma magnetic surfaces. Also possible is a change of the ripple modelling, the current exponential model having only limited accuracy on the high field side of the machine.

## 6. Conclusion

Up to now, Tokaflu has been a very lively model. The model is integrated in a finite element code Castem 2000 and is very practical to use, without laborious data manipulation. The coding also allows easy modification. First validations were made by confrontation to previous heat patterns, and further are being done with the current PFCs of Tore Supra. The code has proved very useful to make progress in our understanding of the heat deposition principles. Tokaflu associated with finite element modelling was extensively used in the design of the PFCs for the CIEL project, to validate bonding technology, optimise components shapes, calculate the power reaching secondary components and help the design of diagnostics.